\documentclass[5p,letter]{elsarticle}

\usepackage{lineno,hyperref}

\journal{Nucl. Instrum. Meth. A}

\begin{document}

\begin{frontmatter}

\title{
Reflectance dependence of polytetrafluoroethylene on thickness for xenon scintillation light}


\author[UMICH]{J.~Haefner}
\author[UMICH]{A.~Neff}
\author[UMICH]{M.~Arthurs}
\author[UMICH]{E.~Batista}
\author[UMICH]{D.~Morton}
\author[UMICH]{M.~Okunawo}
\author[UMICH]{K.~Pushkin}
\author[UMICH]{A.~Sander}
\author[UMICH,UCDAVIS]{S.~Stephenson}
\author[UMICH]{Y.~Wang}
\author[UMICH]{W.~Lorenzon\corref{mycorrespondingauthor}}
\cortext[mycorrespondingauthor]{Corresponding author}

\address[UMICH]{Randall Laboratory of Physics, University of Michigan, Ann Arbor, Michigan 48109-1040, U.S.A.}
\address[UCDAVIS]{University of California Davis, Department of Physics, One Shields Ave., Davis, California 95616, U.S.A.}

\begin{abstract}
Many rare event searches including dark matter direct detection and neutrinoless double beta decay experiments take advantage of the high VUV reflective surfaces made from polytetrafluoroethylene (PTFE) reflector materials to achieve high light collection efficiency in their detectors. As the detectors have grown in size over the past decade, there has also been an increased need for ever thinner detector walls without significant loss in reflectance to reduce dead volumes around active noble liquids, outgassing, and potential backgrounds. We report on the experimental results to measure the dependence of the reflectance on thickness of two PTFE samples at wavelengths near 178\,nm. No change in reflectance was observed as the wall thickness of a cylindrically shaped PTFE vessel immersed in liquid xenon was varied between 1\,mm and 9.5\,mm.
\end{abstract}


\end{frontmatter}


\section{Introduction}
\label{sec:introduction}

Liquid xenon (LXe) detectors have found many applications based on their capability to provide both calorimetry and imaging of particle interactions. Particularly fruitful applications include dark matter~\cite{XENON100-08,ZEPLINIII,LUX,PandaX} and lepton flavor violating~\cite{MEG} searches, neutrinoless double beta decay detectors~\cite{EXO}, gamma-ray physics experiments~\cite{LXeGRIT}, medical imaging~\cite{med-image}, and neutron detection for Homeland Security~\cite{HLS}.

The performance of these detectors is strongly affected by their xenon scintillation light collection efficiency, which depends significantly on the reflectance of the surfaces that surround the LXe volumes. Polytetrafluoroethylene (PTFE) reflector materials are oftentimes the material of choice for LXe detectors. PTFE is known to be highly reflective in the visible and near infrared (NIR) regions~\cite{Weidner-1981}, where it can reach  reflectance of $\mathcal{O}$({99\%}). Reflectance remains high even in the vacuum ultraviolet (VUV) region. For xenon scintillation light ($\lambda\simeq178$\,nm) PTFE reflectance of  $\mathcal{O}$({55\%}) has been measured at room temperature~\cite{Silva-2010,Silva201059}. When immersed in LXe, PTFE reflectance is of $\mathcal{O}$({97\%})~\cite{Akerib:2012ys,Neves:2016tcw}, which is much higher than the $\mathcal{O}$({90\%}) expected from optical models of the PTFE--LXe surface derived from vacuum measurements with xenon scintillation~\cite{Silva-PhD}.

It is important to know the magnitude of PTFE reflectance in LXe, since a few percent difference in PTFE reflectance has a noticeable impact on the performance of a LXe detector~\cite{LZ-CDR}. Absolute PTFE reflectance in LXe has been measured recently~\cite{Neves:2016tcw}, but little is known about the PTFE reflectance in LXe as a function of PTFE thickness at these wavelengths~\cite{Tsai:2008}. The need for studying the impact of thickness is driven by the desire to minimize the amount of PTFE reflector materials in rare event searches while maintaining good reflectance and thus high light collection efficiency. Minimizing PTFE thickness is desirable in order to reduce dead volumes around active LXe, outgassing, and potential backgrounds. A lower limit on the thickness is established by the PTFE transmittance to xenon scintillation light, and the need for optical isolation between active and passive regions in LXe detectors.

In this work, experimental results of the reflectance of PTFE immersed in LXe at wavelengths near 178\,nm are reported as the thickness of the PTFE material is varied between 1\,mm and 9.5\,mm. The basic features of the experimental apparatus are described, followed by a discussion of the relative reflectance measurements of two PTFE materials as a function of PTFE thickness. Determination of absolute reflectance of these PTFE samples is beyond the scope of this paper. It has been performed elsewhere~\cite{Neves:2016tcw}.

\section{Apparatus}
\label{sec:apparatus}

The experimental procedure to measure PTFE reflectance is based on a simple approach~\cite{AHLEN1977513,Salamon1979147} to measure reflectivity of painted surfaces in light integration boxes.\footnote{Note that Refs.~\cite{AHLEN1977513} and~\cite{Salamon1979147} have small errors in their ``light  collection efficiency'' formulae, which slightly alter the results in their publications, but do not significantly alter the general behavior.} The basic idea is to measure photon detection efficiency $f$ while varying the fractional area $F$ covered by a photosensitive detector in a closed chamber, such that
\begin{equation}\label{Eq:detection-eff}
  f = \frac{Ft}{Ft+a(1-F)},
  \label{eq:fracarea}
\end{equation}
where $t$ is the probability of a photon to be absorbed by the PMT (ie. detection), and $a$ is the probability of a photon to be absorbed by the PTFE (for details see Ref.~\cite{Haefner2016}). Fractional area is defined as the ratio of the photosensitive area over the total surface area of the closed chamber. Equation~\ref{eq:fracarea} shows that for fixed  $a$ and $t$, photon detection efficiency decreases as fractional area decreases (or as the surface area of the closed chamber increases). Although this simple model does not account for Rayleigh scattering and light absorption in LXe, and therefore does not allow to extract the absolute reflectance of the two PTFE materials, it illustrates that the method is very sensitive to changes in reflectance for highly reflective materials, as discussed in~\ref{app:sims}.

In order to perform the reflectance measurements, a chamber was designed and built that  a) contains reflecting surfaces made of PTFE material, b) contains a photomultiplier tube (PMT) that is sensitive in the VUV region, c) contains a bright mono-energetic light source inside the detector, d) provides a way to easily modify the fractional area of the chamber, and d) allows variation of the PTFE wall thickness without modifying the experimental condition inside the chamber. Good optical coupling between the PMT window and LXe is obtained, since the refractive index of LXe for intrinsic scintillation light ($n = 1.67$~\cite{Solovov-2004}) and that of quartz (n = 1.57) are well matched.

\begin{figure}[htbp]
\centering
\includegraphics[width=0.7\linewidth]{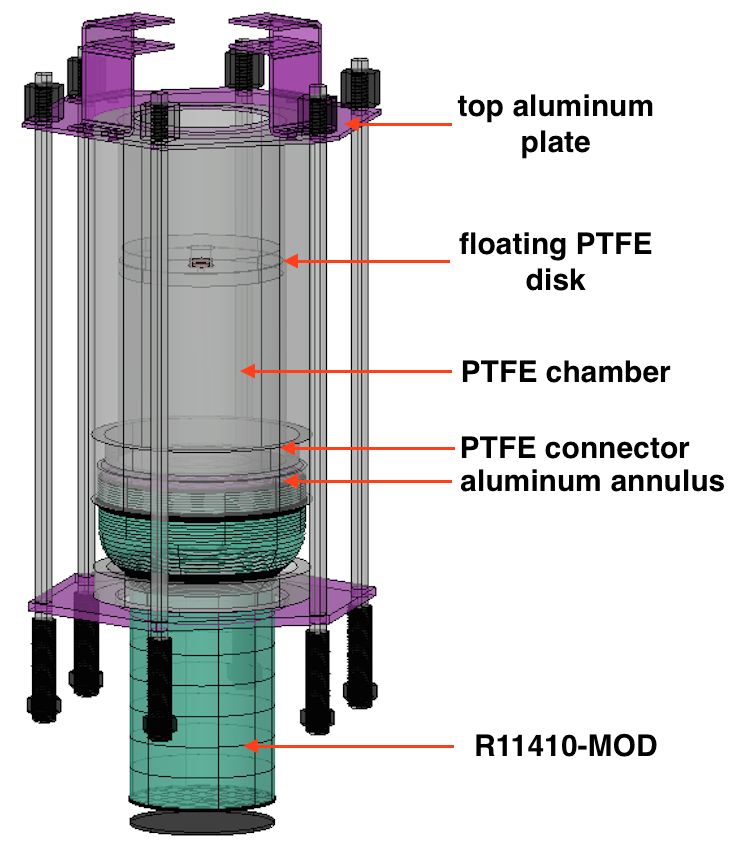}
\hspace*{3mm}
\caption{Schematic view of the reflectivity setup. It consists of a 3-inch Hamamatsu R11410-MOD PMT which covers the bottom part of the 117-mm high and 62.2-mm ID PTFE cylindrical chamber, and a $^{210}$Po source, imbedded in a PTFE floating disk, to provide a mono-energetic light source.
}
\label{fig:setup}
\end{figure}

\begin{figure}[htbp]
\centering
\includegraphics[width=0.65\linewidth]{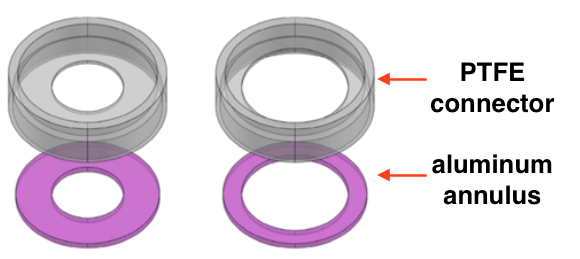} \hspace*{3mm}
\caption{Visualization of the two geometries that allow to vary the effective PMT area, showing the PTFE connectors and the aluminum annuli. Geometry 1 (left) has a 36.8\,mm I.D. for both pieces, while geometry 2 (right) has a 62.2\,mm I.D. for both pieces which is the same as the I.D. of the cylindrical PTFE chamber.}
\label{fig:geometry}
\end{figure}

\begin{figure}[tb]
\centering
\includegraphics[width=0.8\linewidth]{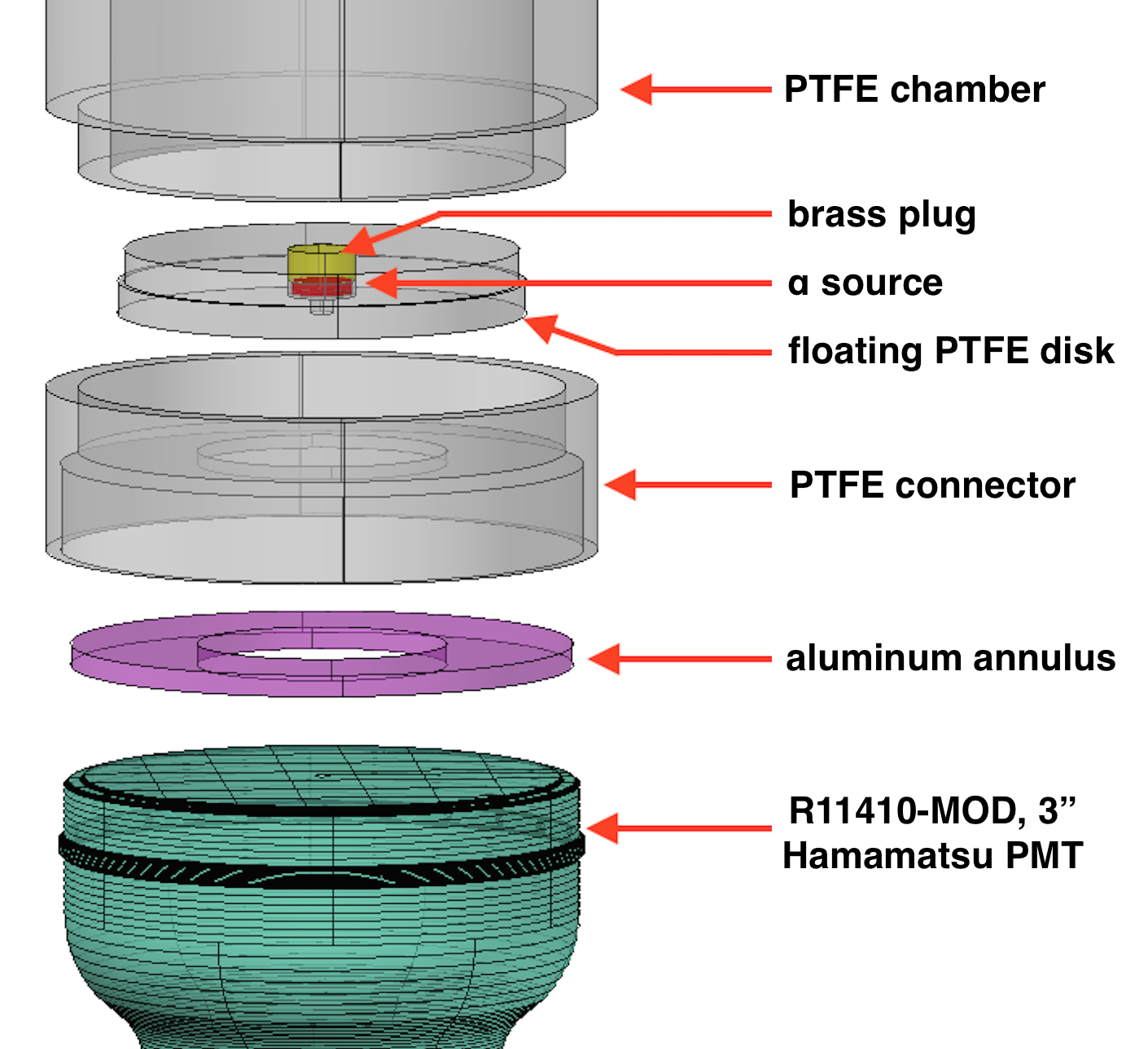} \hspace*{3mm}
\caption{Exploded view of the reflectivity setup for configuration 1. The $^{210}$Po $\alpha$ source, depicted as the small red disk inside the floating disk, is held in place by a brass plug which is secured by a thin sheet of PTFE (not shown) from the backside of the floating disk.
}
\label{fig:exploded}
\end{figure}

A schematic illustration of the detector is shown in Fig.~\ref{fig:setup}. It consists of a cylindrical PTFE chamber (62.2\,mm I.D., 117\,mm high) with a 3-inch Hamamatsu R11410-MOD PMT~\cite{Akerib:2012da}  on the bottom, a circular PTFE disk (60\,mm diameter) that floats on LXe, and a $^{210}$Po source (0.1\,$\mu$Ci) that is chemically plated onto a silver disk and attached to the floating PTFE disk facing down. Two different geometries were used to vary the exposed area of the PMT, as shown in Fig.~\ref{fig:geometry}. In geometry 1, the PMT area was defined by a 3.3\,mm thick and 36.8-mm I.D. aluminum annulus placed on the PMT to protect it from light transmitted through the 2.0\,mm thick aperture of the PTFE connector above it with the same I.D. For geometry 2, the aluminum annulus and the PTFE connector had the same I.D. as the cylindrical PTFE chamber. An exploded view of the detector in which the placement of the aluminum annulus, the PTFE connector, and the surrounding pieces for geometry 1 are emphasized, is presented in Fig.~\ref{fig:exploded}.

Before the chamber is filled, the floating PTFE disk rests on the PTFE connector (for geometry 1) or on the PMT window (for geometry 2). As the chamber is filled with LXe, the floating disk rises and exposes an increasing fraction of the cylindrical PTFE wall, until it makes contact with the top aluminum plate. The top aluminum plate constrains the floating disk even when the level of LXe exceeds the top of the chamber. A 3.2-mm diameter hole in the bottom of the floating disk provides a near point-like light source of about 1,000 cps. Mono-energetic scintillation light is generated by the 5.304\,MeV $\alpha$ particles emitted from $^{210}$Po through  ionization and excitation of xenon atoms and subsequent emission of 178\,nm photons from atomic de-excitation and recombination. Thus, as the floating disk rises,  the number of light reflections on the PTFE surface of the chamber increases and the corresponding path for the VUV photons gets longer, resulting in a smaller light yield,\footnote{Photon detection efficiency depends  both on the properties of the liquid (ie. absorption length and Rayleigh scattering) and on PTFE reflectance.} as illustrated in Sec.~\ref{sec:results}.

The thickness of the PTFE cylinder wall was modified (from 9.5\,mm down to 1\,mm) by removing  material from the outside of the cylinder. This ensured that the inner surface of the chamber remained unchanged throughout the measurements. The minimum cylinder wall thickness of 1\,mm was chosen to avoid collapse of the PTFE cylinder, as there is considerable buoyant force on the PTFE cylinder due to the PMT being immersed in LXe. The thickness of the floating disk and the PTFE connector remained unchanged to avoid modification of the apparent intensity of the light source placed in the floating disk and to maintain structural integrity of the PTFE chamber, respectively.

The detector described here is integrated into the Michigan Xenon (MiX) detector system which is documented in detail elsewhere~\cite{MiX-2015}. The detector shares all components with MiX, except for data acquisition, since only a single PMT is read out. Instead of using a flash ADC, the PMT signal is read out directly on a digital oscilloscope (Tektronix TDS 3034B).

\section{Results}
\label{sec:results}

To prepare for data collection, the PTFE samples were cleaned for 30 minutes in an ultrasonic bath with a solution of alconox and de-ionized water at 50\,C, followed by a 30 minute ultrasonic de-ionized water rinse at 50\,C before being placed in a dry nitrogen box overnight. Prior to measurements the PTFE samples were kept in vacuum ($10^{-3}$\,mbar) for about 24 hours. This is a standard procedure to clean PTFE samples for use in ultrapure LXe detectors.

At the start of a measurement cycle, the PTFE chamber was filled with LXe until the floating disk reached the top aluminum  plate. The xenon was subsequently purified~\cite{MiX-2015} with a high-temperature zirconium SAES \mbox{PS3-MT3-R-1} getter for about 24 hours. Periodic light yield data were recorded at a LXe temperature of $(175\pm0.2)$\,K. Between measurements, the LXe level, measured using a concentric cylindrical capacitor~\cite{MiX-2015} was reduced in steps of about 3\,pF, which corresponded to a change in LXe level of about 8.1 mm. This process was continued until the floating disk rested either on the aperture of the PTFE connector at the bottom of the PTFE chamber (for geometry 1) or on the PMT window (for geometry 2). Data was collected for two different types of PTFE materials from Applied Plastics Technology (APT), 807NX and NXT85, at a variety of thicknesses of the cylindrical chamber wall. For 807NX PTFE, data was collected for two different exposed PMT areas to check for consistency among two different chamber geometries, while for NXT85 PTFE, data was only collected for the larger exposed PMT area. In order to avoid PMT saturation, the detector was operated at different high voltages for the three different experimental configurations.

\begin{figure}[tb]
\centering
\includegraphics[width=0.85\linewidth]{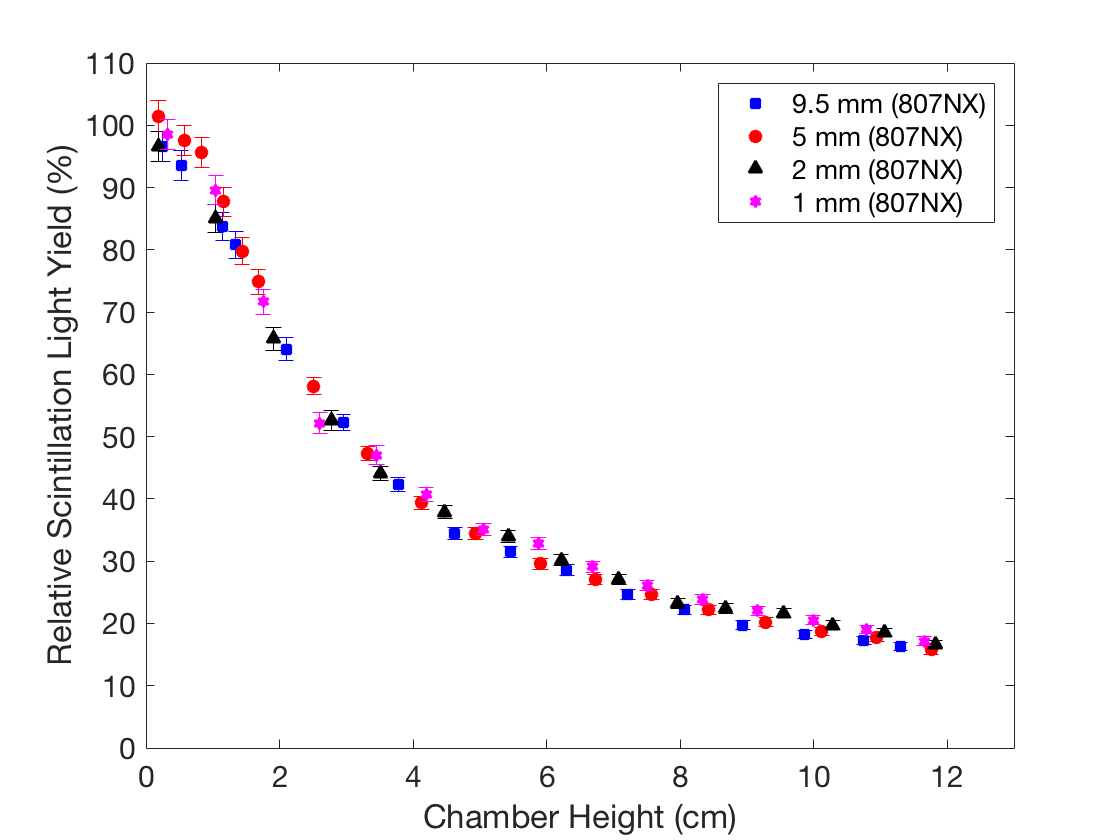} \hspace*{3mm}
\caption{Scintillation light yield versus chamber height for APT 807NX PTFE, scaled to 100\% at the PMT surface. Data were collected at an exposed PMT diameter of 36.8\,mm (geometry 1), a PMT high voltage of 1,150\,V, and chamber wall thicknesses of 9.5\,mm, 5\,mm, 2\,mm, and 1\,mm. The error bars represent the sum of statistical and systematic uncertainties added in quadrature.}
\label{fig:data_807NX_36}
\end{figure}

The data for 807NX PTFE at an exposed PMT diameter of 36.8\,mm are shown in Fig.~\ref{fig:data_807NX_36}. All four data sets, scaled to roughly 100\% at the PMT surface,\footnote{The four data sets were scaled  with the same factor to roughly 100\% at the PMT surface to facilitate reading the fractional decrease in  light yield vs chamber height off the graph.} display the same general shape and agree well within uncertainties. Each data set shows that the scintillation light yield diminishes as the light source moves away from the PMT surface. This general behavior is expected since the scintillation light produced from the 5.304\,MeV $\alpha$ particles will experience more wall bounces at larger distance, resulting in more photons absorbed on the chamber walls. Furthermore, the path length for the scintillation photons increases, resulting in an even smaller light yield due to  photon absorption and Rayleigh scattering in LXe~\cite{Solovov-2004}. The combined effect leads to a variation in scintillation light yield of about a factor of 5.8 over the chamber height of 117\,mm. Note that if reflectance was lower for smaller PTFE wall thickness, the light yield would diminish more rapidly with increasing chamber height, leading to even larger variations over the chamber height.

\begin{figure}[tb]
\centering
\includegraphics[width=0.85\linewidth]{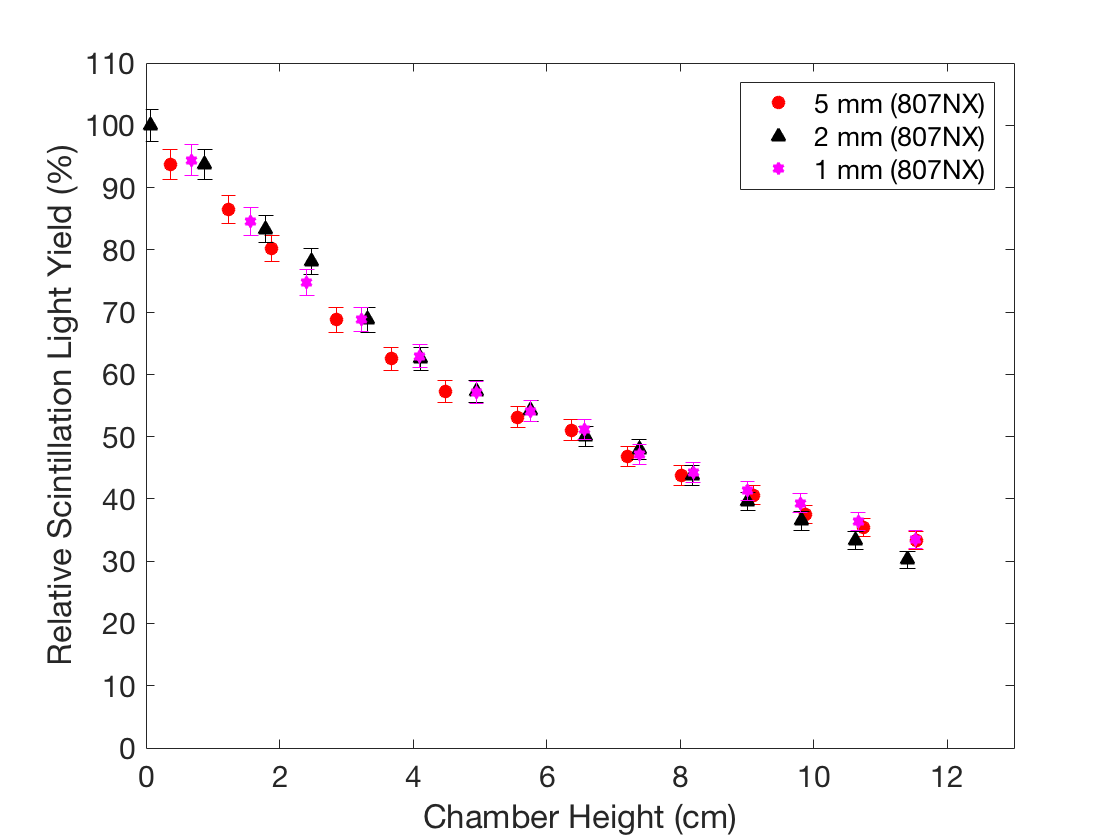} \hspace*{3mm}
\caption{Scintillation light yield versus chamber height for APT 807NX PTFE, scaled to 100\% at the PMT surface. Data were collected at an exposed PMT diameter of 62.2\,mm (geometry 2), a PMT high voltage of 1,010\,V, and chamber wall thicknesses of  5\,mm, 2\,mm, and 1\,mm. The error bars represent the sum of statistical and systematic uncertainties added in quadrature.}
\label{fig:data_807NX_62}
\end{figure}

The data for 807NX PTFE at an exposed PMT diameter of 62.2\,mm are shown in Fig.~\ref{fig:data_807NX_62}. The data sets at 5\,mm and 1\,mm agree very well within uncertainties, but the data set at 2\,mm displays a slightly lower light yield at large chamber height. This variation in light yield is, however, still  consistent within combined statistical and systematic uncertainties and is not necessarily an indication of lower reflectance for a 2\,mm wall thickness. The variation in scintillation light yield over the chamber height is only about a factor of 3 for this configuration, which is in agreement with the expectation that the different chamber geometry results in fewer wall bounces and a shorter path length for scintillation photons to reach the PMT.

\begin{figure}[tb]
\centering
\includegraphics[width=0.85\linewidth]{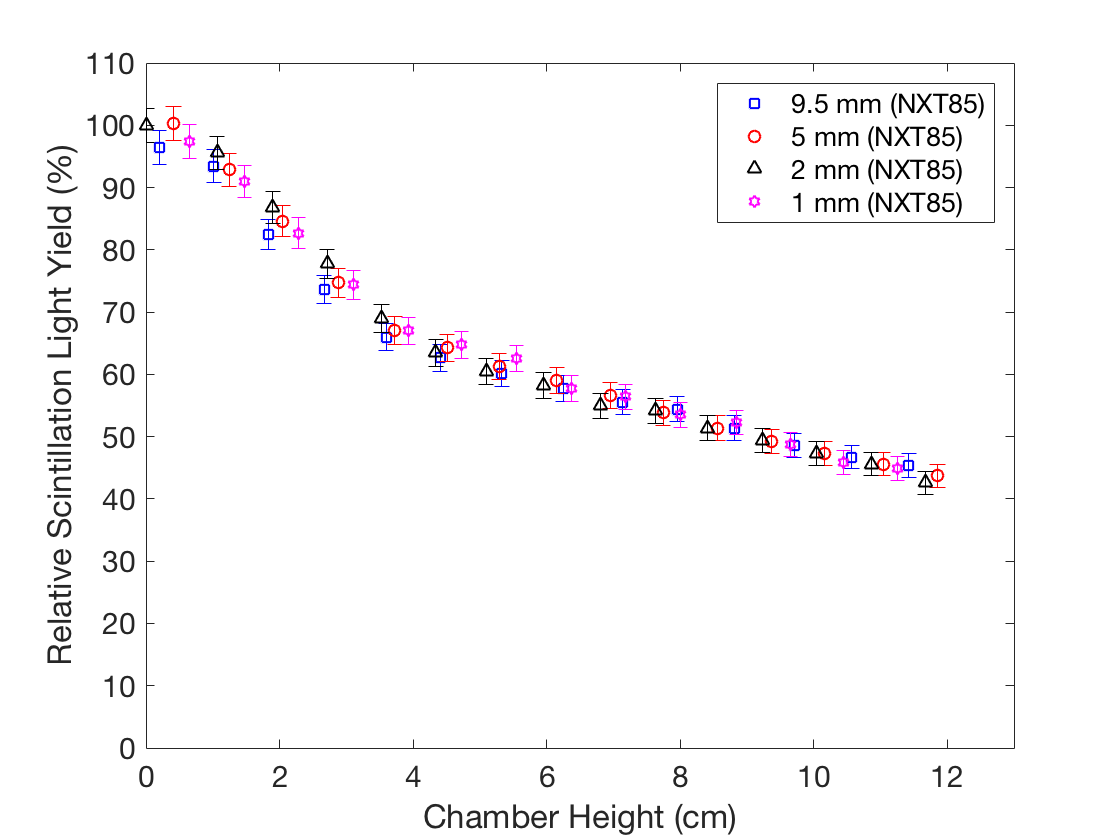} \hspace*{3mm}
\caption{Scintillation light yield versus chamber height, scaled to 100\% at the PMT surface. Data were collected with APT NXT85 PTFE at a PMT diameter of 62.2\,mm  (geometry 2), a PMT high voltage of 930\,V, and chamber wall thicknesses of 9.5\,mm, 5\,mm, 2\,mm, and 1\,mm. The error bars represent the sum of statistical and systematic uncertainties added in quadrature.}
\label{fig:data_NXT85_62}
\end{figure}

The two data sets for the two different chamber geometries for 807NX PTFE show distinctly different shapes, but show no decrease in the signal over the measured chamber wall thicknesses. Therefore, decreasing the wall thickness from 9.5\,mm to 1\,mm does not appear to change the reflectance.

The data for NXT85 PTFE at an exposed PMT diameter of 62.2\,mm are shown in Fig.~\ref{fig:data_NXT85_62}. Note that data were only collected for one exposed PMT area since no reflectance dependence on thickness was observed for the two geometries measured with the 807NX sample. All four data sets show the same general shape and agree well within uncertainties. The variation in scintillation light yield over the chamber height is only about a factor of 2.3 for this configuration.

\begin{figure}[tb]
\centering
\includegraphics[width=0.85\linewidth]{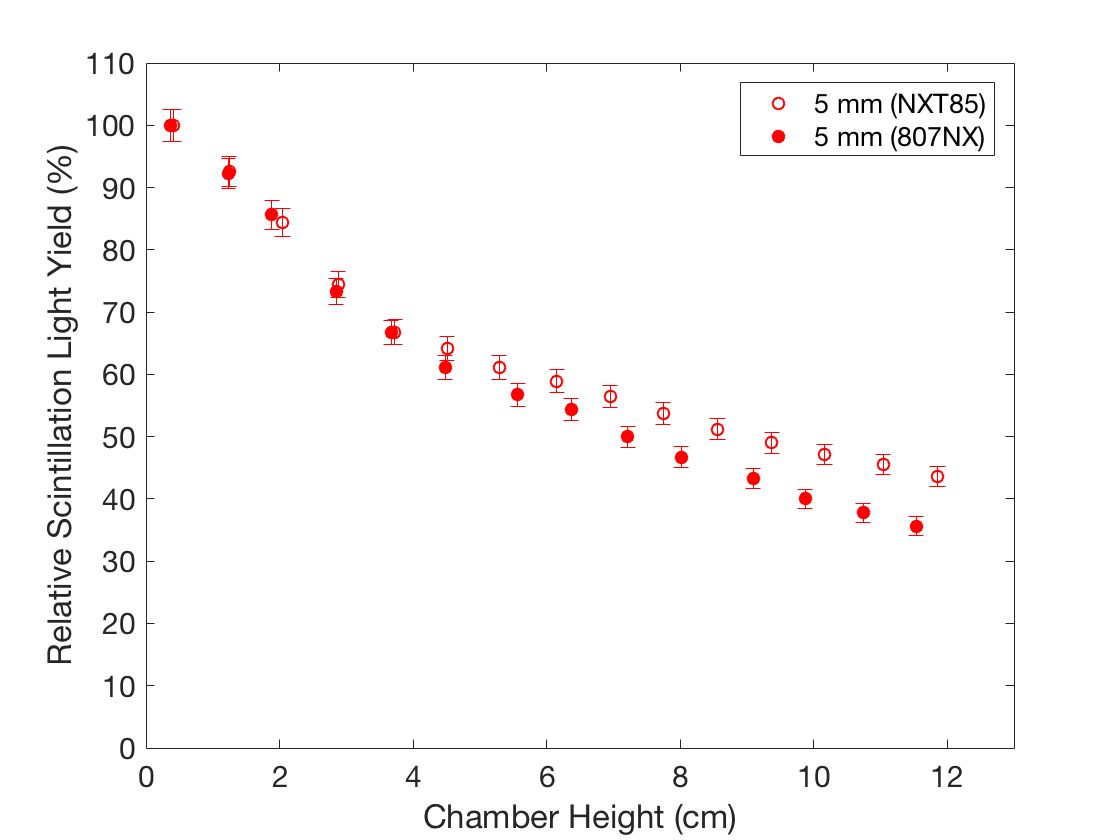} \hspace*{3mm}
\caption{Comparison of scintillation light yield versus chamber height for NXT85 and 807NX PTFE at an exposed PMT diameter of 62.2\,mm (geometry 2). The two data sets were taken with a 5\,mm chamber wall thickness, and normalized to unity at the PMT surface. The error bars represent the sum of statistical and systematic uncertainties added in quadrature.}
\label{fig:data_both}
\end{figure}

Figure~\ref{fig:data_both} shows a direct comparison of scintillation light yield as a function of chamber height for the two PTFE materials. Both data sets were collected with exposed PMT diameters of 62.2\,mm and a cylinder wall thickness of 5\,mm. The two data sets were normalized to unity at the PMT surface to extract the fractional decrease in light yield from minimum exposed PTFE area to maximum exposed PTFE area. The fractional decrease will be smaller when the reflectance is larger.
As shown in Fig.~\ref{fig:data_both}, the NXT85 data set displays a higher light yield at large chamber height than the 807NX data set. This indicates that NXT85 PTFE has reflectance that is slightly higher than that of 807NX PTFE, in agreement with the results obtained for the absolute reflectance measured from Ref.~\cite{Neves:2016tcw}. See ~\ref{app:estimates} for a discussion of the sensitivity of the measurements reported in Figs.~\ref{fig:data_807NX_36}, \ref{fig:data_807NX_62}, \ref{fig:data_NXT85_62} to variations of reflectance as a function of PTFE thickness.

\subsection{Temperature Dependence and Experimental Uncertainties}
\label{sec:tempdep}

\begin{figure}[tbp]
\centering
\includegraphics[width=0.85\linewidth]{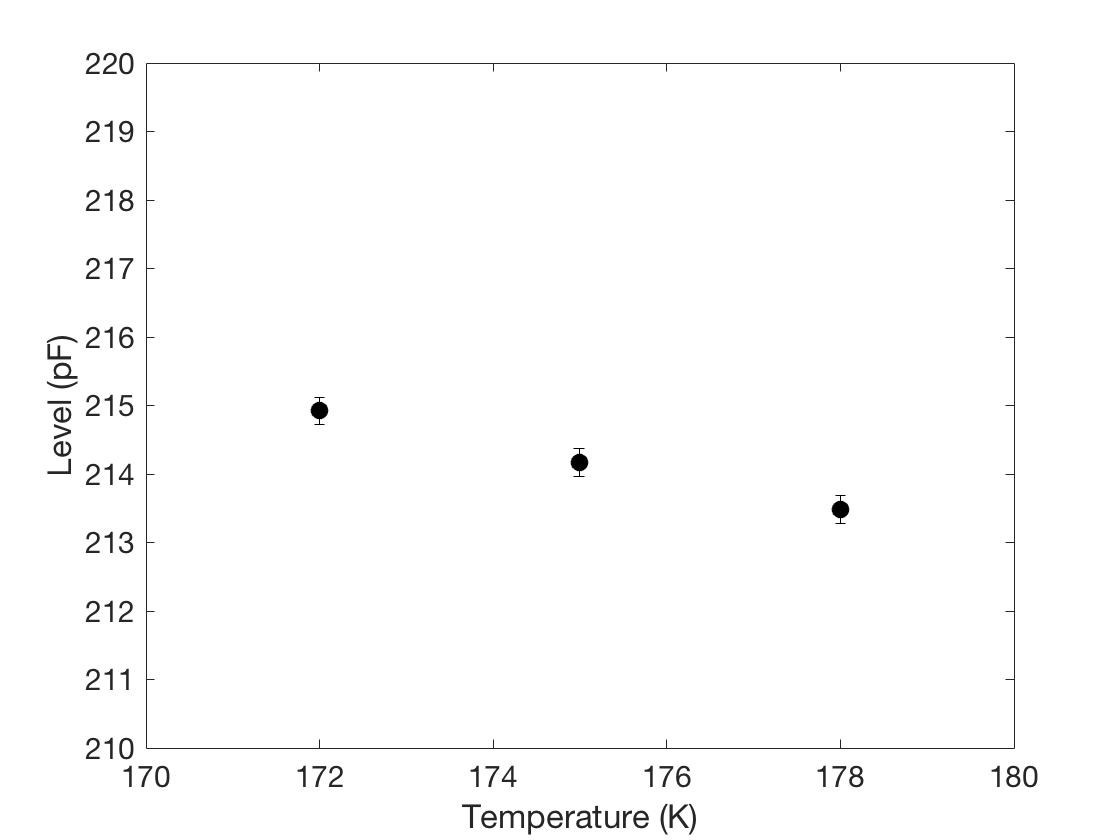} \hspace*{3mm}
\caption{Level meter reading at various temperatures with reflectance measurement setup at a PMT diameter of 62.2\,mm. The level meter was fully submerged in LXe at all temperatures. The data were collected with the 807NX PTFE sample.}
\label{fig:lev_v_temp}
\end{figure}

\begin{figure}[tbp]
\centering
\includegraphics[width=0.85\linewidth]{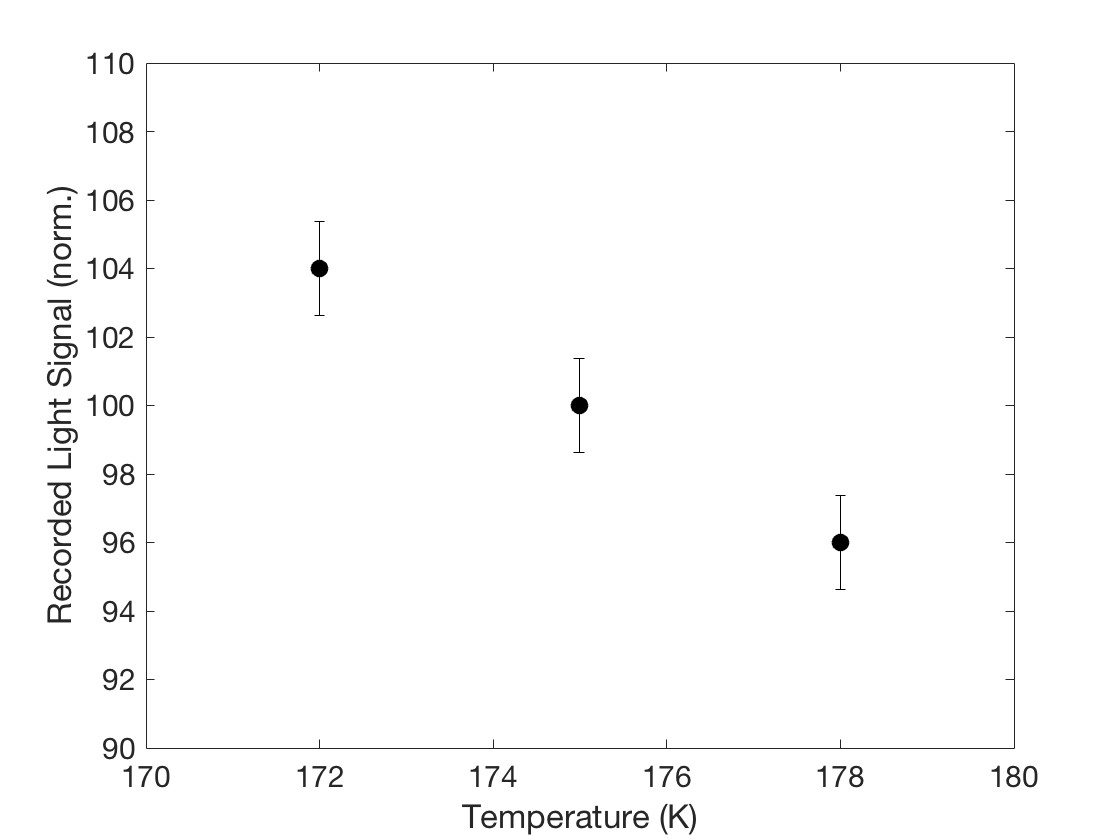} \hspace*{3mm}
\caption{Scintillation light yield at various temperatures with a PMT diameter of 62.2\,mm, normalized to unity at 175\,K. PMT held at 1010\,V. Source location fixed. The data were collected with the 807NX PTFE sample.}
\label{fig:sig_v_temp}
\end{figure}

In order to quantify the temperature dependence of the capacitive level meter and of the scintillation light signal recorded in the PMT, the temperature of the LXe was varied within $\pm$3\,K of the nominal temperature of 175\,K. The measurements were performed with the detector filled beyond the top aluminum plate to ensure that the source location remained fixed, and that the level meter was fully submerged in LXe. The results are shown on Figs.~\ref{fig:lev_v_temp} and \ref{fig:sig_v_temp}. Figure~\ref{fig:lev_v_temp} exhibits a drop in the level meter reading with increasing temperature. This behavior is expected, as an increase in temperature leads to a drop in the dielectric constant of LXe, and therefore to a drop in capacitance of the cylindrical level meter. The temperature dependence of the capacitance is $-0.1\%$/K, in good agreement with literature values~\cite{crc96th}. Figure~\ref{fig:sig_v_temp} displays an increase of scintillation light signal in LXe with decreasing temperature. This temperature dependence is also expected, but its value of $-1.3\%$/K is approximately three times larger than typically reported values~\cite{Sauli1992}. Since the PMT was maintained at the same temperature as LXe, the larger-than-expected variation in the signal may in fact be due to the dependence of the PMT gain on temperature. Control of the LXe temperature at the $\pm0.2$\,K level has been routinely achieved in the detector, corresponding to a lower than 0.3\% variation of scintillation light yield, and lower than 0.03\% variation of the LXe level.

Capacitive level meter calibration was achieved at the $\pm0.2$\,pF (ie. $\pm0.54$\,mm LXe) level, which translates into a variation in the scintillation light yield of about $\pm2$\%. Each measurement cycle was repeated at least once, typically within 48 hours, by refilling the chamber with LXe and letting it settle for at least 24 hours before starting the next measurement cycle. Scintillation light yield was found to be stable within typically $\pm1$\% between two consecutive measurement cycles. The total systematic uncertainty was estimated to be about $\pm2.3$\% by adding the individual  instrumental uncertainties in quadrature.

\section{Conclusions}
\label{sec:conclusions}

In summary, an experimental study has been performed to determine the change in reflectance of PTFE immersed in LXe at $\lambda=178$\,nm  as the thickness of the PTFE material was reduced from 9.5\,mm down to 1\,mm. The data, collected with two different PTFE samples, displayed no reduction in reflectance down to 1\,mm wall thickness. The ability of PTFE to maintain good reflectance and thus high light collection efficiency even at very thin thicknesses is instrumental for next generation direct detection dark matter search and neutrinoless double beta decay experiments. These experiments require large size detectors, small dead space, low outgassing, minimal potential radioactive backgrounds, and good optical isolation between active and passive regions within these detectors, requirements that can all be met by thin sheets of PTFE.

No attempt was made to extract the absolute reflectivity of the two PTFE samples, as this is beyond the scope of this paper. Nevertheless, the data indicate that the reflectance of NXT85 PTFE is slightly higher than the reflectance of 807NX PTFE for xenon scintillation light.

\appendix
\section{Sensitivity to Changes in Reflectance}
\label{app:sims}

\begin{figure}[tb]
\centering
\includegraphics[width=0.49\linewidth]{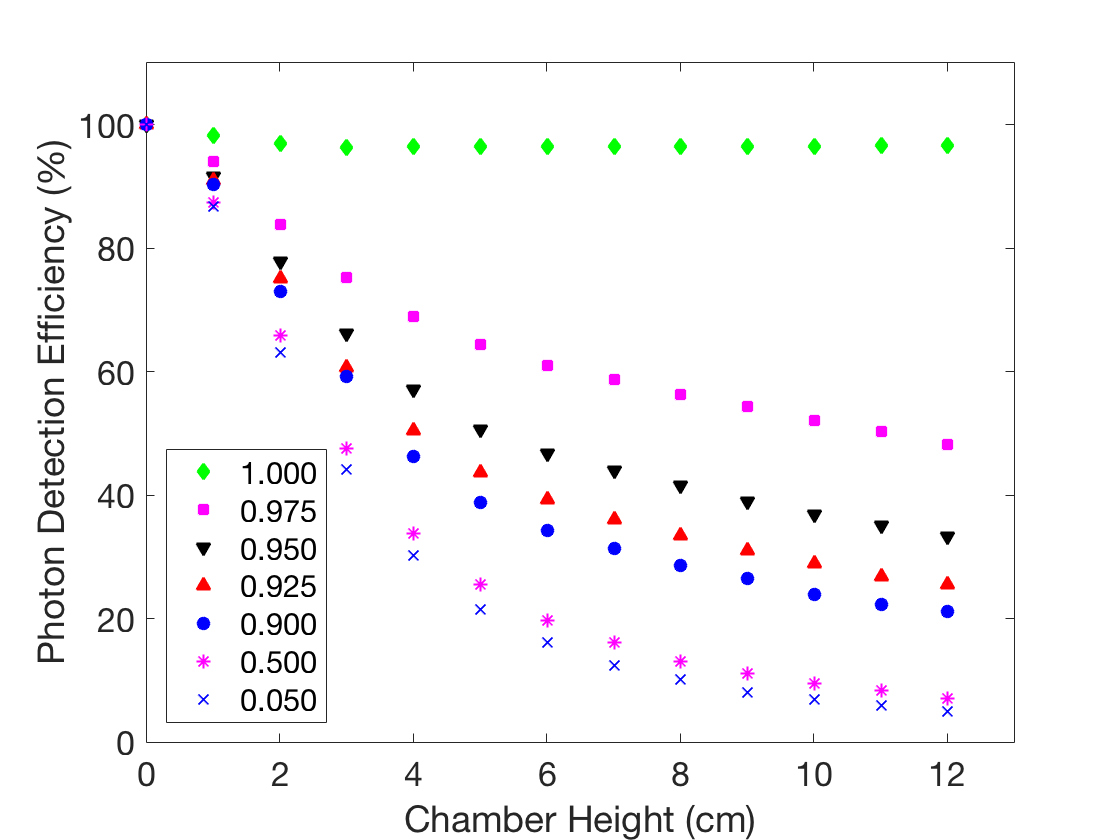}
\includegraphics[width=0.49\linewidth]{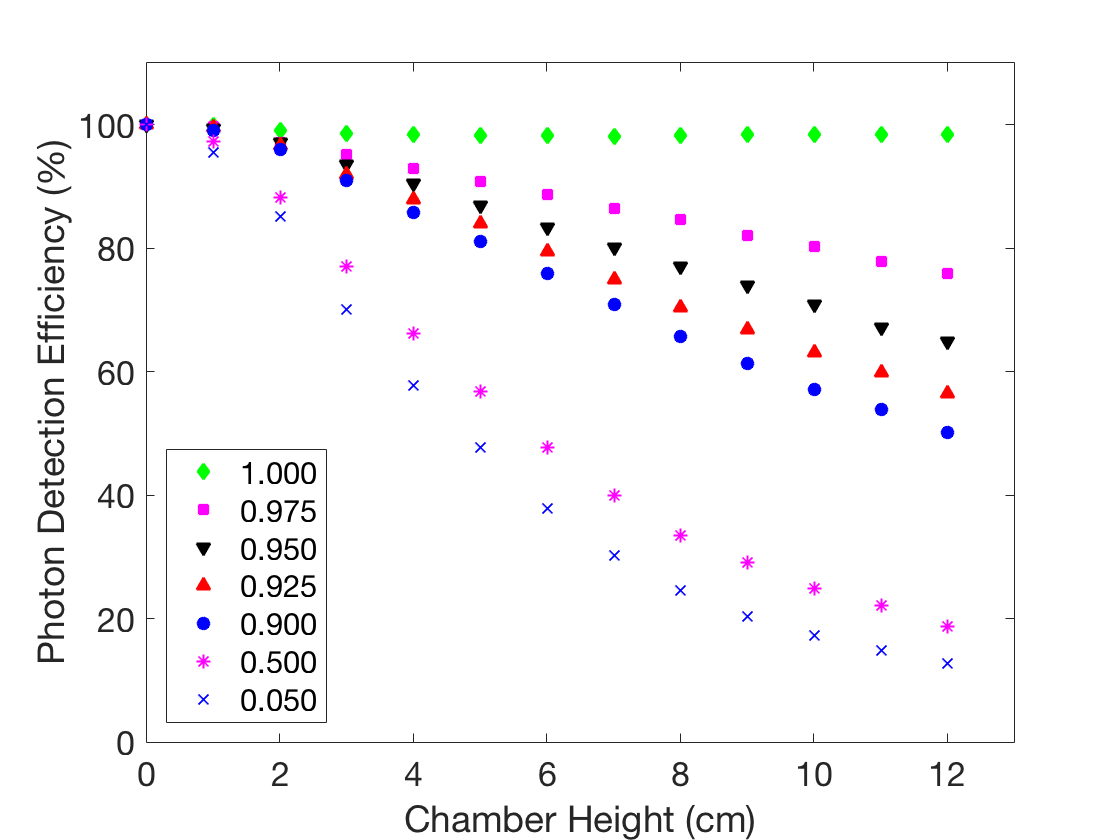}
\hspace*{-1mm}
\caption{Simulations of scintillation light yield versus chamber height for PTFE reflectances between 90\% and 100\% for geometry 1 (left) and geometry 2 (right). Note that the simulations do not account for Rayleigh scattering and light absorption in LXe, but they do account for the effect of light absorption on the aluminum annulus, as reflected by the small drop in photon detection efficiency below unity for a PTFE reflectance of 100\%.
}
\label{fig:sims}
\end{figure}

Simulations of the experimental setup were performed that follow the approach introduced in Refs.~\cite{AHLEN1977513,Salamon1979147,Haefner2016} to illustrate the sensitivity of the method to changes in reflectance. The results of the simulations, shown in Fig.~\ref{fig:sims}, display photon detection efficiency as a function of chamber height, rather than fractional area to facilitate comparison to the experimental data displayed in Figs.~\ref{fig:data_807NX_36}, \ref{fig:data_807NX_62}, \ref{fig:data_NXT85_62}, and \ref{fig:data_both}. The left panel of Fig.~\ref{fig:sims} shows the variation of the photon detection efficiency for geometry 1 over a large range of PTFE reflectances, while the right panel shows the variation for geometry 2. Sensitivity to variations of reflectance is high at high reflectances $\mathcal{O}$({$>$95\%}), gradually diminishes as reflectance gets smaller, and gets poor as reflectance gets low $\mathcal{O}$({$<$50\%}). Since absolute PTFE reflectance is of $\mathcal{O}$({97\%})~\cite{Neves:2016tcw} when immersed in LXe, this method is very sensitive to small variations of reflectance and thus well-suited to detect possibly small variations in reflectance with PTFE thickness.

Note that the general features of light yield versus chamber height are well reproduced for the two different geometries. But the model is too simple to allow an extraction of the absolute reflectance, since it does neither account for Rayleigh scattering and light absorption in LXe, nor does it include detailed models for diffuse and specular reflection. Building a model that can determine the absolute reflectance of the measured PTFE samples is beyond the scope of this paper.

\section{Estimation of Variation in Reflectance}
\label{app:estimates}

The sensitivity of the measurements reported in Figs.~\ref{fig:data_807NX_36}, \ref{fig:data_807NX_62}, \ref{fig:data_NXT85_62} to variations of reflectance as a function of PTFE thickness can be quantified from examining the results displayed in Fig.~\ref{fig:data_both}. As shown in the figure, the NXT85 data set displays a higher light yield at large chamber height than the 807NX data set, which is in agreement with the results obtained for the absolute reflectance measured from Ref.~\cite{Neves:2016tcw} who report values of 97.5\% and 96.1\% for NXT85 and 807NX PTFE, respectively. Combining the 1.4\% difference in absolute reflectance of the two PTFE samples with the difference in light yield at large chamber height displayed in Fig.~\ref{fig:data_both}, the variations of reflectance versus PTFE wall thickness for the three experimental configurations were estimated. The fluctuations of the data shown in Figs.~\ref{fig:data_807NX_36}, \ref{fig:data_807NX_62}, and \ref{fig:data_NXT85_62} near chamber height of 12\,cm are were determined to be consistent with fluctuations in reflectance of 0.4\% for configuration 1 (Fig.~\ref{fig:data_807NX_36}), of 0.5\% for configuration 2 (Fig.~\ref{fig:data_807NX_62}), and of 0.4\% for configurations 3 (Fig.~\ref{fig:data_NXT85_62}), respectively.

\section*{Acknowledgments}

We thank Carl Akerlof and Richard Raymond for many useful discussions, and Harry Nelson and Vladimir Solovov for careful reading of the manuscript. We would also like to acknowledge the LZ collaboration for providing the PTFE materials. This work was supported by DOE grants DE-SC0015708 and DE-SC0010830, and by generous funds from the University of Michigan.

\section*{References}

\bibliography{PTFEbibfile_v2}

\begin{thebibliography}{10}
\expandafter\ifx\csname url\endcsname\relax
  \def\url#1{\texttt{#1}}\fi
\expandafter\ifx\csname urlprefix\endcsname\relax\def\urlprefix{URL }\fi
\expandafter\ifx\csname href\endcsname\relax
  \def\href#1#2{#2} \def\path#1{#1}\fi

\bibitem{XENON100-08}
J.~Angle, \textit{et al.}, {First Results from the XENON10 Dark Matter
  Experiment at the Gran Sasso National Laboratory}, Phys. Rev. Lett. 100
  (2008) 021303.
\newblock \href {http://dx.doi.org/10.1103/PhysRevLett.100.021303}
  {\path{doi:10.1103/PhysRevLett.100.021303}}.

\bibitem{ZEPLINIII}
V.~Lebedenko, \textit{et al.}, {Result from the First Science Run of the
  ZEPLIN-III Dark Matter Search Experiment}, Phys. Rev. D 80 (2009) 052010.
\newblock \href {http://dx.doi.org/10.1103/PhysRevD.80.052010}
  {\path{doi:10.1103/PhysRevD.80.052010}}.

\bibitem{LUX}
D.~Akerib, \textit{et al.} [LUX~Collaboration], {Results from a Search for Dark
  Matter in the Complete LUX Exposure}, Phys. Rev. Lett. 118 (2017) 021303.
\newblock \href {http://dx.doi.org/10.1103/PhysRevLett.118.021303}
  {\path{doi:10.1103/PhysRevLett.118.021303}}.

\bibitem{PandaX}
A.~Tan, \textit{et al.} [PandaX-II~Collaboration], {Dark Matter Results from
  First 98.7 Days of Data from the PandaX-II Experiment}, Phys. Rev. Lett. 117
  (2016) 121303.
\newblock \href {http://dx.doi.org/10.1103/PhysRevLett.117.121303}
  {\path{doi:10.1103/PhysRevLett.117.121303}}.

\bibitem{MEG}
S.~Ritt, \textit{et al.} [MEG~Collaboration], {Status of the MEG expriment},
  Nuc. Phys. B Supp. 162 (2006) 279--282.
\newblock \href {http://dx.doi.org/10.1016/j.nuclphysbps.2006.09.091}
  {\path{doi:10.1016/j.nuclphysbps.2006.09.091}}.

\bibitem{EXO}
M.~Auger, \textit{et al.} [EXO~Collaboration], {Search for Neutrinoless
  Double-Beta Decay in $^{136}$Xe with EXO-200}, Phys. Rev. Lett. 109 (2012)
  032505.
\newblock \href {http://dx.doi.org/10.1103/PhysRevLett.109.032505}
  {\path{doi:10.1103/PhysRevLett.109.032505}}.

\bibitem{LXeGRIT}
A.~Curioni, E.~Aprile, T.~Doke, K.~Giboni, M.~Kobayashi, U.~Oberlack, {A study
  of the LXeGRIT detection efficiency for MeV gamma-rays during the 2000
  balloon flight campaign}, Nuc. Instrum. Meth. 576 (2007) 350--361.
\newblock \href {http://dx.doi.org/10.1016/j.nima.2007.02.102}
  {\path{doi:10.1016/j.nima.2007.02.102}}.

\bibitem{med-image}
H.~Zaklad, S.~Derenzo, R.~A. Muller, G.~Smadja, R.~G. Smits, L.~W. Alvarez, {A
  Liquid Xenon Radioisotope Camera}, IEEE Trans. Nucl. Sci. 19 (1972) 206--213.
\newblock \href {http://dx.doi.org/10.1109/TNS.1972.4326727}
  {\path{doi:10.1109/TNS.1972.4326727}}.

\bibitem{HLS}
J.~Nikkel, T.~Gozani, C.~Brown, J.~Kwong, D.~McKinsey, Y.~Shin, S.~Kane,
  C.~Gary, M.~Firestone, {Liquefied Noble Gas (LNG) detectors for detection of
  nuclear materials}, JINST 7 (2012) 206--213.
\newblock \href {http://dx.doi.org/10.1088/1748-0221/7/03/C03007}
  {\path{doi:10.1088/1748-0221/7/03/C03007}}.

\bibitem{Weidner-1981}
V.~Weidner, J.~Hsia, Reflection properties of pressed polytetrafluoroethylene
  powder, J. Opt. Soc. Am. 71 (1981) C03007.
\newblock \href {http://dx.doi.org/10.1364/JOSA.71.000856}
  {\path{doi:10.1364/JOSA.71.000856}}.

\bibitem{Silva-2010}
C.~Silva, J.~P. da~Cunha, A.~Pereira, V.~Chepel, M.~I. Lopes, V.~Solovov,
  F.~Neves, Reflectance of polytetrafluoroethylene for xenon scintillation
  light, J. Appl. Phys. 107 (2010) 064902.
\newblock \href {http://dx.doi.org/10.1063/1.3318681}
  {\path{doi:10.1063/1.3318681}}.

\bibitem{Silva201059}
C.~Silva, J.~P. da~Cunha, A.~Pereira, M.~Lopes, V.~Chepel, V.~Solovov,
  F.~Neves, A model of the reflection distribution in the vacuum ultra violet
  region, Nucl. Instrum. Meth. A619~(1–3) (2010) 59 -- 62, frontiers in
  radiation physics and applications: Proceedings of the 11th International
  Symposium on Radiation Physics.
\newblock \href {http://dx.doi.org/10.1016/j.nima.2009.10.086}
  {\path{doi:10.1016/j.nima.2009.10.086}}.

\bibitem{Akerib:2012ys}
D.~Akerib, \textit{et al.} [LUX~Collaboration], {The Large Underground Xenon
  (LUX) Experiment}, Nucl. Instrum. Meth. A704 (2013) 111--126.
\newblock \href {http://arxiv.org/abs/1211.3788} {\path{arXiv:1211.3788}},
  \href {http://dx.doi.org/10.1016/j.nima.2012.11.135}
  {\path{doi:10.1016/j.nima.2012.11.135}}.

\bibitem{Neves:2016tcw}
F.~Neves, A.~Lindote, A.~Morozov, V.~Solovov, C.~Silva, P.~Bras, J.~P.
  Rodrigues, M.~I. Lopes, {Measurement of the absolute reflectance of
  polytetrafluoroethylene (PTFE) immersed in liquid xenon, }\href
  {http://arxiv.org/abs/1612.07965} {\path{arXiv:1612.07965}}.

\bibitem{Silva-PhD}
C.~da~Silva, Study of the reflectance distributions of fluoropolymers and other
  rough surfaces with interest to scintillation detectors, Ph.D. thesis,
  University of Coimbra (2009).

\bibitem{LZ-CDR}
D.~Akerib, \textit{et al.} [LZ~Collaboration], {LUX-ZEPLIN (LZ) Conceptual
  Design Report, }\href {http://arxiv.org/abs/1509.02910}
  {\path{arXiv:1509.02910}}.

\bibitem{Tsai:2008}
B.~K. Tsai, D.~W. Allen, L.~M. Hanssen, B.~Wilthan, J.~Zeng, {A comparison of
  optical properties between solid PTFE (Teflon) and (low density) sintered
  PTFE} (2008).

\bibitem{AHLEN1977513}
S.~Ahlen, B.~Cartwright, G.~Tarl\'{e}, Technique for painting light integration
  boxes, Nucl. Instrum. Meth. 143~(3) (1977) 513 -- 517.
\newblock \href {http://dx.doi.org/10.1016/0029-554X(77)90240-3}
  {\path{doi:10.1016/0029-554X(77)90240-3}}.

\bibitem{Salamon1979147}
M.~Salamon, G.~Tarl\'{e}, An algorithm to calculate light distributions within
  light diffusion boxes, Nuclear Instruments and Methods 161~(1) (1979) 147 --
  150.
\newblock \href {http://dx.doi.org/10.1016/0029-554X(79)90373-2}
  {\path{doi:10.1016/0029-554X(79)90373-2}}.

\bibitem{Haefner2016}
J.~Haefner, {Measurements and Analysis of Polytetrafluoroethylene Reflectance
  at the Liquid Xenon Scintillation Radiation Wavelength}, Honors thesis, Ann
  Arbor, Michigan, USA (2016).

\bibitem{Solovov-2004}
V.~N. Solovov, V.~Chepel, M.~I. Lopes, A.~Hitachi, R.~{Ferreira Marques}, A.~J.
  P.~L. Policarpo, {Measurement of the refractive index and attenuation length
  of liquid xenon for its scintillation light}, Nucl. Instrum. Meth. A516
  (2004) 462--474.
\newblock \href {http://dx.doi.org/10.1016/j.nima.2003.08.117}
  {\path{doi:10.1016/j.nima.2003.08.117}}.

\bibitem{Akerib:2012da}
D.~S. Akerib, et~al., {An Ultra-Low Background PMT for Liquid Xenon Detectors},
  Nucl. Instrum. Meth. A703 (2013) 1--6.
\newblock \href {http://arxiv.org/abs/1205.2272} {\path{arXiv:1205.2272}},
  \href {http://dx.doi.org/10.1016/j.nima.2012.11.020}
  {\path{doi:10.1016/j.nima.2012.11.020}}.

\bibitem{MiX-2015}
S.~Stephenson, J.~Haefner, Q.~Lin, K.~Ni, K.~Pushkin, R.~Raymond, M.~Schubnell,
  N.~Shutty, G.~Tarl\'{e}, C.~Weaverdyck, W.~Lorenzon, \mbox{MiX}: a position
  sensitive dual-phase liquid xenon detector, JINST 10~(10) (2015) P10040.
\newblock \href {http://dx.doi.org/10.1088/1748-0221/10/10/P10040}
  {\path{doi:10.1088/1748-0221/10/10/P10040}}.

\bibitem{crc96th}
W.~Haynes, \href{https://books.google.com/books?id=RpLYCQAAQBAJ}{CRC Handbook
  of Chemistry and Physics, 96th Edition}, CRC Press, 2015.
\newline\urlprefix\url{https://books.google.com/books?id=RpLYCQAAQBAJ}

\bibitem{Sauli1992}
F.~Sauli, \href{https://books.google.com/books?id=4HVEorltBeIC}{Instrumentation
  in High Energy Physics}, Advanced series on directions in high energy
  physics, World Scientific, 1992.
\newline\urlprefix\url{https://books.google.com/books?id=4HVEorltBeIC}

\end{thebibliography}

\end{document}